# Modelling Silica using MACE-MP-0 Machine Learnt Interatomic Potentials

Jamal Abdul Nasir, Jingcheng Guan, Woongkyu Jee, Scott M. Woodley, Alexey A. Sokol, C. Richard A. Catlow, Alin-Marin Elena*

[1] Jamal Abdul Nasir, Jingcheng Guan, Woongkyu Jee, Scott M. Woodley, Alexey A. Sokol, C. Richard A. Catlow

Department of Chemistry, Kathleen Lonsdale Materials Chemistry

University College London, 20   Gordon Street, London WC1H 0AJ, United Kingdom

[2] C. Richard A. Catlow

UK Catalysis Hub, Research Complex at Harwell,

 Rutherford Appleton Laboratory, R92 Harwell, Oxfordshire OX11 0FA, United Kingdom

[3] C. Richard A. Catlow

 School of Chemistry, Cardiff University

  Park Place, Cardiff, CF10 3AT, United Kingdom

[4] Alin-Marin Elena

  STFC Scientific Computing Department, Daresbury Laboratory,

     Keckwick Lane, Daresbury, Warrington, WA4 4AD, United Kingdom.

 E-mail: alin-marin.elena@stfc.ac.uk

**ABSTRACT**. Silica polymorphs and zeolites are fundamental to a wide range of industrial applications owing to their diverse structural characteristics, thermodynamic and mechanical stability under varying conditions and due to their geological importance. Computational modelling has played a crucial role in understanding the relationship between the structure and functionality of silicas and silicates including zeolites. In this study, we apply the MACE-MP-0 machine learnt interatomic potentials (ML-IP) to model the framework energies of siliceous zeolites and examine the phase transitions of silica and ZSM-5 polymorphs under high-pressure conditions. The results reproduce the known metastability of siliceous zeolites relative to α-quartz, with energy differences between microporous and dense phases calculated by MACE-MP-0 medium ML-IP and density functional theory (DFT) methods closely aligning with experimental calorimetric data. The high-pressure simulations reveal distinct compression behaviour in the quartz, coesite, and stishovite polymorphs of silica, with coesite and stishovite showing increased stability at elevated pressures in line with experimental data. The calculated phase transition pressures from quartz to coesite (~3.5 GPa) and coesite to stishovite (~9 GPa) are close to experimental findings, demonstrating the reliability of MACE ML-IP in modelling the structural and energetic properties of silica polymorphs.

# Introduction

Dense silicas and silicates are intensively studied materials owing to their geological and industrial importance. Their microporous counterparts, zeolites, including both aluminosilicate and silica materials, have numerous industrial applications, including catalysis, gas adsorption, and ion exchange, due to their highly diverse tuneable chemical and structural properties[1]. Classical interatomic potential (IP) based techniques and Density Functional Theory (DFT) have been widely and successfully applied to modelling both dense and microporous silicas and silicates [2-4]. Machine learning (ML) techniques[5, 6], offer new opportunities but their viability in modelling these materials accurately has not been explored in detail. However, recent developments in foundational potentials and advanced parameterization techniques of ML potentials have begun to address these challenges, improving their accuracy and reliability in modelling these materials[7, 8]. While specific MLIP for zeolites exist[9-11] in this work we assess the suitability of the off-the-shelf MLIP, MACE-MP-0 method to model zeolites.

Structure enumeration techniques have identified more than two million possible zeolite frameworks [12-14], but only 240 zeolite frameworks have been synthesized and listed in the International Zeolite Association (IZA) database [15]. This discrepancy is often referred to as the "zeolite conundrum" [14]. As a result, ongoing research is focused on advanced synthesis methods,[6-9] It is well known that microporous materials are metastable compared to their dense polymorphs [16, 17], and some useful correlations between the energies of siliceous zeolites relative to α-quartz and their framework densities have been already established [18, 19]. In addition, computational methods play a crucial role in the discovery of new zeolite materials by enabling the exploration and classification of both known and hypothetical structures [13, 20, 21]. Hence, we have selected the problem of relative structural stability of main dense and several known microporous silicas as the first test of the MACE ML potentials.

Further, in tests of the viability of new energy landscape methods applied to materials, an appealing problem to consider is pressure-driven phase transitions. Considering dense silicas, the high-pressure transition from quartz to coesite and then from coesite to stishovite has long been of significant interest in geophysics and geochemistry, and consequently, the physical properties and stability relations of these three polymorphs have been extensively studied[22-24]. Several experimental investigations have aimed to determine the precise transition boundaries between quartz and coesite, as well as between coesite and stishovite since accurate measurements of these transitions can serve as important pressure standards at high temperatures [25, 26]. Additionally, the elastic properties of quartz, coesite, and stishovite have been examined using various experimental techniques. Among microporous materials, notably, ZSM-5 zeolites including their purely siliceous form, silicalite, show polymorphism[27], crystallizing in an orthorhombic (Pnma)[28], monoclinic (P21/n11)[29], and orthorhombic (P212121)[30] lattice undergoing low-to-high symmetry transitions with temperature or pressure, and here we will concentrate on the former two structures known as silicalite-1. Further, in tests of the viability of new energy landscape methods applied to materials, an appealing problem to consider is pressure-driven phase transitions. Considering dense silicas, the high-pressure transition from quartz to coesite and then from coesite to stishovite has long been of significant interest in geophysics and geochemistry, and consequently, the physical properties and stability relations of these three polymorphs have been extensively studied[22-24]. Several experimental investigations have aimed to determine the precise transition boundaries between quartz and coesite, as well as between coesite and stishovite since accurate measurements of these transitions can serve as important pressure standards at high temperatures [25, 26]. Additionally, the elastic properties of quartz, coesite, and stishovite have been examined using various experimental techniques. Among microporous materials, notably, ZSM-5 zeolites including their purely siliceous form, silicalite, show polymorphism, crystallizing in an orthorhombic (Pnma)[28], monoclinic (P21/n11)[29], and orthorhombic (P212121)[30] lattice undergoing low-to-high symmetry transitions with temperature or pressure, and here we will concentrate on the former two structures known as silicalite-1. Traditional IP methods have been applied to study these phase transformations[31] which can be used as a useful guide.

On the phase transition from coesite to stishovite the silicon changes its coordination number from four to six. A much less well-characterised but important in the field of microporous silicates is an increase in the coordination number from four to five in some of the high silica framework materials synthesised following the fluoride route [32, 33]. The fluoride ion in open channels or pores readily attaches to one of the framework silicon ions but remains stable in central regions of smaller cages, e.g., double four-membered rings (D4R) where silicon ions retain their original coordination [34]. Reproduction of this behaviour for the systems where it has earlier been observed experimentally and/or studied with first-principles calculations thus forms the last suitable challenge we will consider.

In this study, we thus investigate the performance of MACE a graph neural Message Passing machine learnt interatomic potential which includes Atomic Cluster Expansion in modelling the framework stability of several siliceous materials in their dense and microporous forms followed by a study of the phase transitions of the ground state dense silica polymorph — quartz first to coesite and then coesite to stishovite under pressure, as well as another pressure-driven phase transition of a microporous silica polymorph—silicalite-1 from monoclinic to orthorhombic. To test further the range of applicability of the ML potential, we apply it to the case of fluoride-modified zeolites. Our MACE-ML IP results closely match those predicted by DFT techniques and reported from the experiment, achieving very good chemical accuracy.

## Methodology

### Machine-Learned Interatomic Potential Model: MACE

We have used the MACE architecture, ML-IP framework designed for atomistic simulations[35-37]. The MACE architecture is grounded in equivariant message-passing graph tensor networks that retain key geometrical and physical symmetries of atomic structures, making it highly suitable for simulating diverse materials and chemical processes[36]. This model was chosen for its accuracy in capturing the potential energy surfaces (PES) of materials while ensuring computational efficiency comparable to classical force fields. MACE builds on the Atomic Cluster Expansion (ACE) approach[38], employing higher body-order equivariant features. In this implementation, we used a model with four-body equivariant features and two layers of message passing to capture complex atomic interactions. The radial cutoff was set at 6Å, with the function of interatomic distances expanded into 10 Bessel functions[39], which was followed by a smooth polynomial cutoff function to construct radial features that were fed into a fully connected feed-forward neural network. The training process involved the use of the publicly available MPtrj dataset[40], which contains over 1.5 million configurations of inorganic crystals, derived from DFT geometry optimisations using the Perdew–Burke–Ernzerhof (PBE) functional[41]. To ensure accuracy and efficiency, we selected the medium-sized model ($L=1$) from the MACE framework for all simulations, which balances computational cost and fitting precision. For all our calculations we employed the medium version of MACE-MP-0 in the form of its latest release, so-called 0b-agnesi, in addition, we added empirical D3 correction. All single point and geometry optimisation calculations were done using janus-core code[42]. Geometry optimisation was done employing algorithms implemented in an Atomistic Simulation Environment, specifically L-BFGS and FrechetCellFilter – when cell parameters were optimised. [43]

## Results and Discussion

### Modelling Framework Energies

**Table 1** presents a comparison between our MACE ML-IP calculated relative framework energies with respect to α-quartz (the ground state silica polymorph), and the IP[44], DFT[44] calculated and experimental calorimetric data from Navrotsky and co-workers[17, 45]. MACE_MP_0-ML-IP medium model produces remarkably close results when compared to DFT (PBE+D3), especially across a wide variety of zeolite frameworks. The results show that IP lattice energy methods generally predict higher values than experiment, while DFT and MACE ML-IP results are much closer, within ~1 kJ/mol of each other.

**Table 1**. Calculated and experimental normalised zeolite energies per T-site with respect to α-quartz.

| Structure | IP/kJ/mol | DFT/kJ/mol | ML-IP/kJ/mol | EXP/kJ/mol |
| --- | --- | --- | --- | --- |
| AFI | 11.9 | 10.0 | 10.5 | 7.2 |
| AST | 18.1 | 12.7 | 13.8 | 10.9 |
| BEA | 14.4 | 11.0 | 11.3 | 9.3 |
| CFI/CIT-5 | 12.7 | 12.0 | 12.0 | 8.8 |
| CHA | 16.1 | 12.2 | 12.9 | 11.4 |
| IFR/ITQ-4 | 15.0 | 10.3 | 10.2 | 10 |
| MEL/ZSM-11 | 10.8 | 9.2 | 9.4 | 8.2 |
| MFI/ZSM-5 | 9.7 | 8.3 | 8.5 | 6.8 |
| MWW/ITQ-1 | 14.4 | 11.2 | 11.2 | 10.4 |
| STT/SSZ-23 | 14.7 | 11.4 | 11.4 | 9.2 |
| EMT | 20.1 | 13.0 | 13.3 | 10.5 |
| FER | 11.8 | 9.6 | 10.0 | 6.6 |
| MEI/ZSM-18 | 18.9 | 13.0 | 13.3 | 13.9 |
| α-cristobalite | 3.4 | 5.1 | 2.5 | 2.64 |
| α-tridymite | 4.4 | 6.8 | 3.6 | 5.3 |
| Coesite | 2.02 | 1.85 | 2.2 | 5.1 |
| Stishovite | 133.8 | 39.5 | 37.9 | 49.4 |

Comparison of calculated lattice energies using I.P[44], Shell Model (with Sanders potentials)[46], DFT[44], MACE ML-IP, and experimental values [17, 45, 47].

In most cases, the ML-IP results are nearly indistinguishable from DFT, demonstrating the accuracy and reliability of the machine learning approach in predicting framework energies. For example, the ML-IP energy for AFI (10.5 kJ/mol) closely aligns with the DFT result (10.0 kJ/mol). Similarly, for MFI/ZSM-5, ML-IP predicts an energy of 8.5 kJ/mol, nearly identical to the DFT value of 8.3 kJ/mol. When examining more complex frameworks, such as MEI/ZSM-18 and STT/SSZ-23, ML-IP maintains strong alignment with DFT predictions and the calorimetric data. The ML-IP and DFT energies for STT/SSZ-23 are nearly identical, both at 11.4 kJ/mol, which demonstrates that our ML-IP model is particularly reliable for such structures. For MEI/ZSM-18, the ML-IP value of 13.3 kJ/mol is close to the DFT result of 13.0 kJ/mol, showing a minor deviation but still well within an acceptable range. In general, if better accuracy is needed one can fine-tune MACE_MP, tasks we will explore in future work.

In comparison to the shell model and rigid ion IP calculations, ML-IP consistently outperforms crystal energetics. For example, in the CHA framework, the IP result (16.1 kJ/mol) overestimates the energy compared to both ML-IP (12.9 kJ/mol) and DFT (12.2 kJ/mol), with ML-IP providing a closer match to DFT and experimental values. The same pattern is observed for other frameworks such as MWW/ITQ-1 and BEA, where ML-IP predictions are more aligned with DFT and experimental data than IP results. The strong correlation between ML-IP and DFT highlights the efficiency of the ML-IP method in achieving near-DFT accuracy across different zeolite structures. The few minor discrepancies observed between ML-IP and DFT in complex structures can probably be refined with further fine-tuning of the ML model. The overestimation of the relative crystal energetics of the microporous silicas with respect to the quartz by the IP methods probably arises from the inability of fixed charge potentials to model the effects of small variations in the charge distribution on changing from a dense to a microporous structure, as discussed by Stacey et al [27], although we should note that the IP techniques correctly reproduce trends and also model crystal structures accurately.

## Modelling Phase Transition Energies

Next, we compute the change in energy $\Delta E$ of formation between quartz and cristobalite which is a key thermodynamic property, essential for understanding the phase transformation of Silica as shown in Table 2. The ML-IP computed value of $\Delta E$, 2.5 kJ/mol, falls within the experimental range of 1.88 to 2.64 kJ/mol; Kracek[48] measured a value of 0.63 kcal/mol (2.64 kJ/mol) at room temperature, while Holm[47] reported 0.45 kcal/mol (1.88 kJ/mol) at 970 K. We have found strong agreement between MACE ML-IP results and empirical findings which further supports the validity of the MACE ML-IP.

**Table 2**. Enthalpy of formation changes between quartz and cristobalite

| Temperature (K) | Enthalpy Change (kJ/mol) | Reference |
|---|---|---|
| **298** | 2.64 | [48] |
| **970** | 1.88 | [47] |
| **273** | 2.50 | This work |

## High-Pressure Phase Transitions and Structural Response

A major challenge to any modelling techniques when applied to silica and silicates is to predict the response of the materials to pressure, which we explore in this section.

### *Lattice parameters and angle as a function of pressure in SiO$_2$-polymorphs*

Crystalline forms of silica include α- and β-quartz, α- and β-tridymite, α- and β-cristobalite, coesite, keatite, and stishovite. A phase diagram is shown in **Figure 1**. Except for stishovite, these polymorphs consist of SiO$_4$ tetrahedra with minimal bond length and angle distortion[49]. Under high pressures, silica polymorphs undergo structural transitions[50, 51]. Stishovite, with a rutile structure, has silicon octahedrally coordinated by six oxygen atoms[52]. Denser phases with high density, low compressibility, and high elastic modulus are suggested at even higher pressures. The change of unit cell lattice parameters (*a*, *b*, and *c*) and crystallographic angles (α, β, and γ) with increasing pressure for three different polymorphs of $SiO_2$: quartz, coesite, and stishovite.

The behavior of α-quartz, under high-pressure conditions has been the focus of numerous experimental investigations [53, 54]. To examine the role of high pressure on the structural changes of three phases of $SiO_2$, We conducted MACE-ML-IP simulation and observed that as pressure increases, all three phases of $SiO_2$ — quartz, coesite, and stishovite — show a decrease in their lattice parameters[55], though the extent and nature of the compression vary based on their crystal structures. Quartz, with its hexagonal structure, exhibits significant compression along all axes can be seen in **Figure 1a**. This behaviour is consistent with the hexagonal symmetry of quartz, where the atomic arrangement along the *c*-axis is more compressible[56]. Despite the reduction in lattice dimensions, the crystallographic angles remain constant, with α and β at 90° and γ at 120°. In coesite, a high-pressure polymorph of quartz, the compression is more uniform compared to quartz[57], but with a slightly greater reduction along the *a*-axis (**Figure 1b**). The monoclinic structure of coesite adapts to increasing pressure by compressing in a more balanced manner across all axes, suggesting that the atomic arrangement along the *a*-axis is more susceptible to compression. Similar to quartz, the crystallographic angles in the coesite remain stable under pressure. Stishovite, the densest and highest-pressure polymorph of $SiO_2$, behaves differently. The compression in stishovite is anisotropic with the lattice parameters *a*, *b* and *c*-axis remaining relatively stable [58], which indicates that stishovite's tetragonal structure resists compression along the all-axis, likely due to the tight atomic packing. Like quartz and coesite, the angles in stishovite (α, β, and γ) remain fixed at 90º, preserving the tetragonal symmetry under pressure (**Figure 1c**).

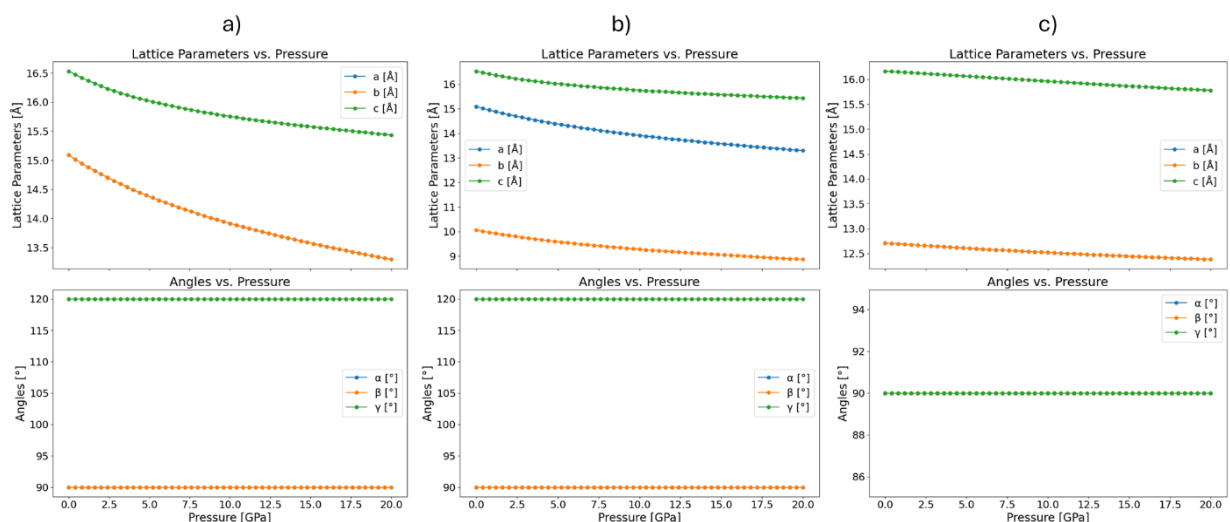

**Figure 1**. Change of the unit cell lattice parameters and angles as a function of pressure (up to 20 GPa) for three different polymorphs of SiO$_2$: a) quartz, b) coesite, and c) stishovite.

*Lattice parameters and angle as a function of pressure in Silicalite polymorphs*

**Figure 2** presents the pressure dependence of lattice parameters and angles for the orthorhombic and monoclinic phases of ZSM-5. In the orthorhombic phase (**Figure 2a**), the lattice parameters *a*, *b*, and *c* exhibit a smooth and gradual decrease with increasing pressure, characteristic of uniform compression. This consistent reduction in lattice dimensions reflects the structural stability of the orthorhombic phase across the entire pressure range, with no evident anomalies. The cell angles remain constant at 90°, indicating the preservation of orthorhombic symmetry throughout the compression process. In contrast, the monoclinic phase (**Figure 2b**) shows a more complex response to increasing pressure. While the lattice parameters also generally decrease with pressure, a noticeable deviation occurs around 1.5 GPa, particularly in the *b* and *c* lattice parameters, suggesting a possible structural reorganization or onset of a phase transition at this pressure.

The most significant difference between the two phases is observed in the behaviour of the β angle. In the monoclinic phase, β increases gradually from 90° to approximately 91.2°, indicating an increase in monoclinic distortion with pressure. However, at ~ 1.5 GPa, there is a sharp drop in the β angle back to 90°, suggesting a transition to a structure with reduced monoclinic distortion or a possible reversion to a higher symmetry phase. This aligns well with the experimental data reported by Mario *et al*.[59], who found that the monoclinic angle α remained unchanged under pressure, indicating no clear tendency toward a transition to the orthorhombic form. We infer, that while the orthorhombic phase remains structurally stable under compression, the monoclinic phase undergoes a significant structural transformation near 1.5 GPa. The pressure-induced change in the β angle in the monoclinic phase suggests a reversible phase transition, possibly involving a shift toward a more stable or symmetric structure at higher pressures.

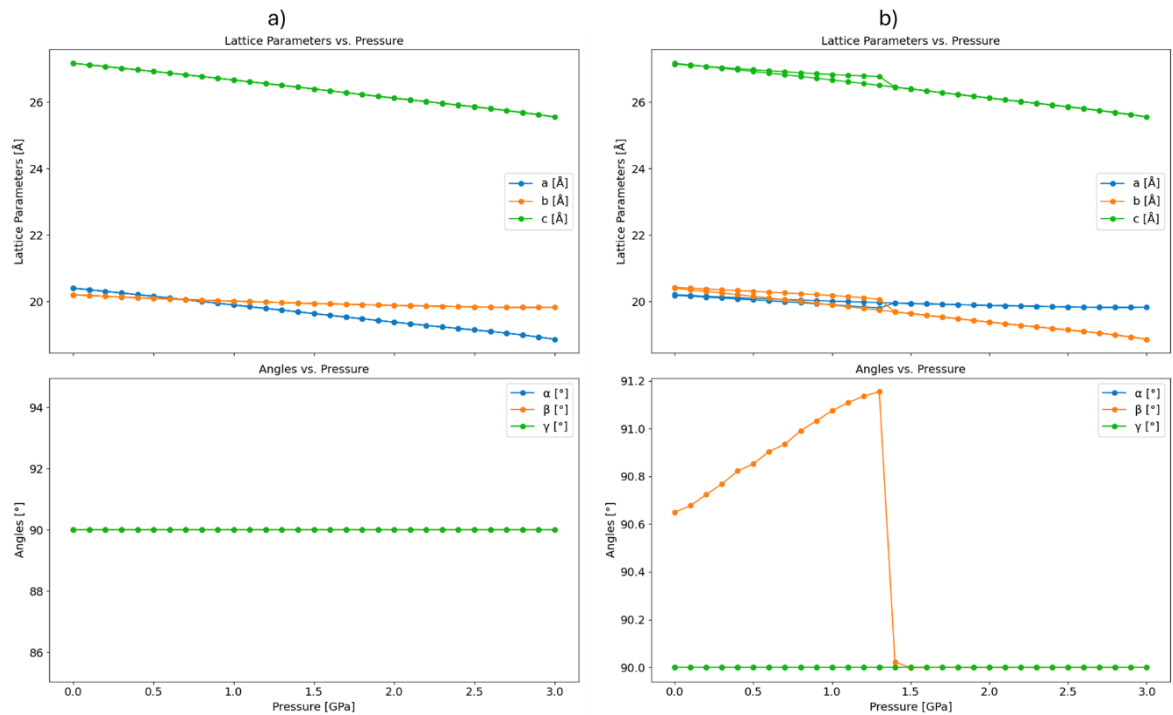

**Figure 2**. Change of the unit cell lattice parameters and angles as a function of pressure (up to 3 GPa) for two different polymorphs of silicalite: a) Orthorhombic, and b) Monoclinic.

### *Volume as a function of pressure in $SiO_2$-polymorphs*

Next, we study the variation in volume as a function of pressure for all three $SiO_2$ polymorphs: quartz, coesite, and stishovite. **Figure 3** illustrates the calculated variation in volume per formula unit with increasing pressure for the three polymorphs. Each of these polymorphs displays distinct behaviour under compression, reflective of their differing crystal structures and stability ranges.

Starting with quartz, which exhibits the most significant reduction in volume with increasing pressure. At ambient conditions, quartz has the largest volume among the three polymorphs [60, 61], which decreases progressively as pressure increases. As observed, the rate of compression is initially rapid, indicating that quartz's hexagonal crystal structure is relatively flexible and can accommodate significant reductions in atomic spacing under pressure.

However, at pressures above 10 GPa, the rate of volume decrease begins to slow, suggesting that the quartz structure becomes increasingly resistant to further compression as it approaches its structural limits. Coesite, however, shows a less pronounced reduction in volume with pressure compared to quartz[62]. The decrease in volume with pressure is more gradual, which is expected given Coesite's formation at relatively higher pressures, where its atomic arrangement is already more compact. Similar to quartz, coesite exhibits a slight reduction in the rate of compression at higher pressures (above 10 GPa), suggesting that increasingly structurally resistant to further volume reduction. Finally, Stishovite, the densest of the three polymorphs[63], demonstrates the least compressibility under pressure, since, its initial volume is significantly lower than both quartz and Coesite and also has minimal volume reduction as pressure increases. The near-flat slope of the volume-pressure curve for stishovite indicates that even at 20 GPa, its structure is quite stable and resistant to further compression, which is in good agreement with *in-situ* synchrotron X-ray diffraction experimental data[64].

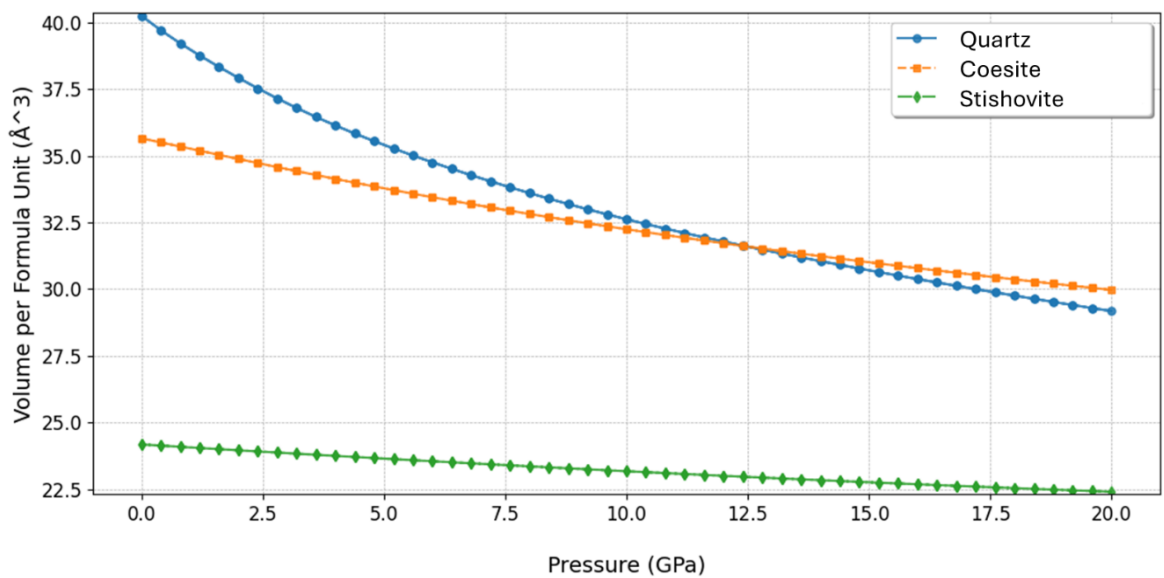

**Figure 3**. Change of unit cell volume (Å³) (per formula unit) as a function of pressure (GPa) for Quartz, Coesite, and Stishovite.

***Volume as a function of pressure in Silicalite polymorphs***

**Figure 4** illustrates the volume behaviour of monoclinic and orthorhombic ZSM-5 polymorphs under varying pressure. Initially, both polymorphs show a linear decrease in volume with increasing pressure up to about 1.0 GPa, with the monoclinic phase displaying a slightly larger volume than the orthorhombic phase. Around 1.35 GPa, the monoclinic phase undergoes a sharp volume reduction, signalling a phase transition. After this point, the volume-pressure behaviour of both phases converges, suggesting similar structural responses to pressure beyond the transition. This phase transition aligns with previously reported pressure ranges for the monoclinic to orthorhombic transformation in ZSM-5, typically occurring between 1.0 and 1.5 GPa[65, 66]. The orthorhombic phase, meanwhile, shows a consistent and smooth decrease in volume, indicating its stability throughout the pressure range.

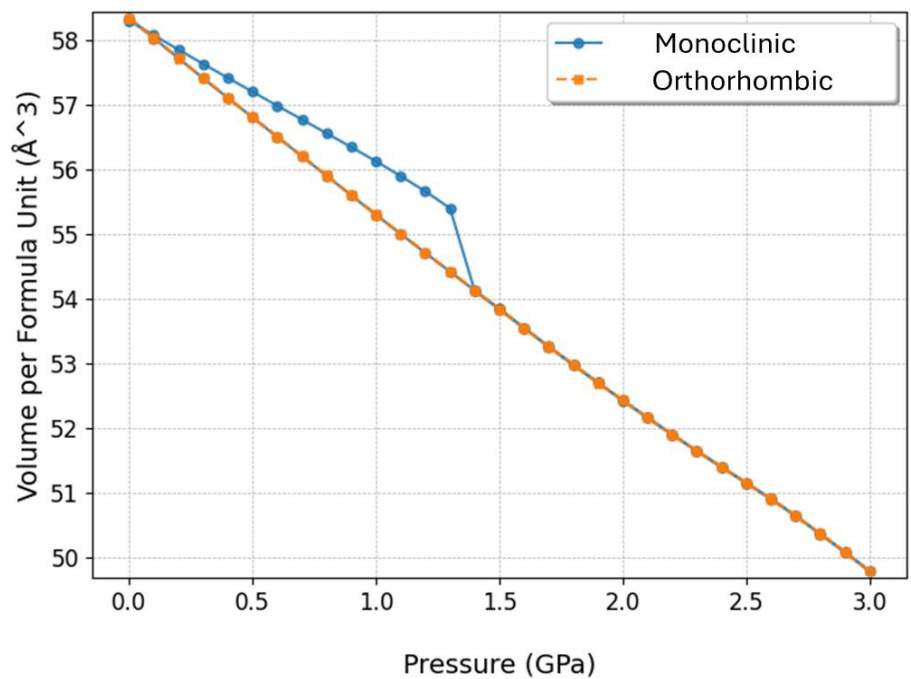

**Figure 4**. Change of unit cell volume (Å³) (per formula unit) as a function of pressure (GPa) for Monoclinic and Orthorhombic polymorphs of ZSM-5.

*Enthalpy as a function of pressure in Dense $SiO_2$ polymorphs*

We have calculated the enthalpy as a function of pressure for the silica polymorphs quartz, Coesite, and Stishovite using MACE ML-IP as shown in **Figure 5** and **Table 3**. Our results show distinct linear relationships between enthalpy and pressure for each phase, which we have correlated with experimental data [47]. The transition from quartz to coesite in our calculations occurs at approximately ~3 GPa, consistent with the experimentally observed transition pressure ~2.5 Gpa [47]. The coesite to stishovite transition, which we calculate occurs at approximately ~9 GPa, shows good agreement with experimental findings (8-8.5 GPa) [47]. Studies reported that Stishovite is stable at pressures between ~9 GPa and 50 GPa[25, 67], which also aligns well with our findings.

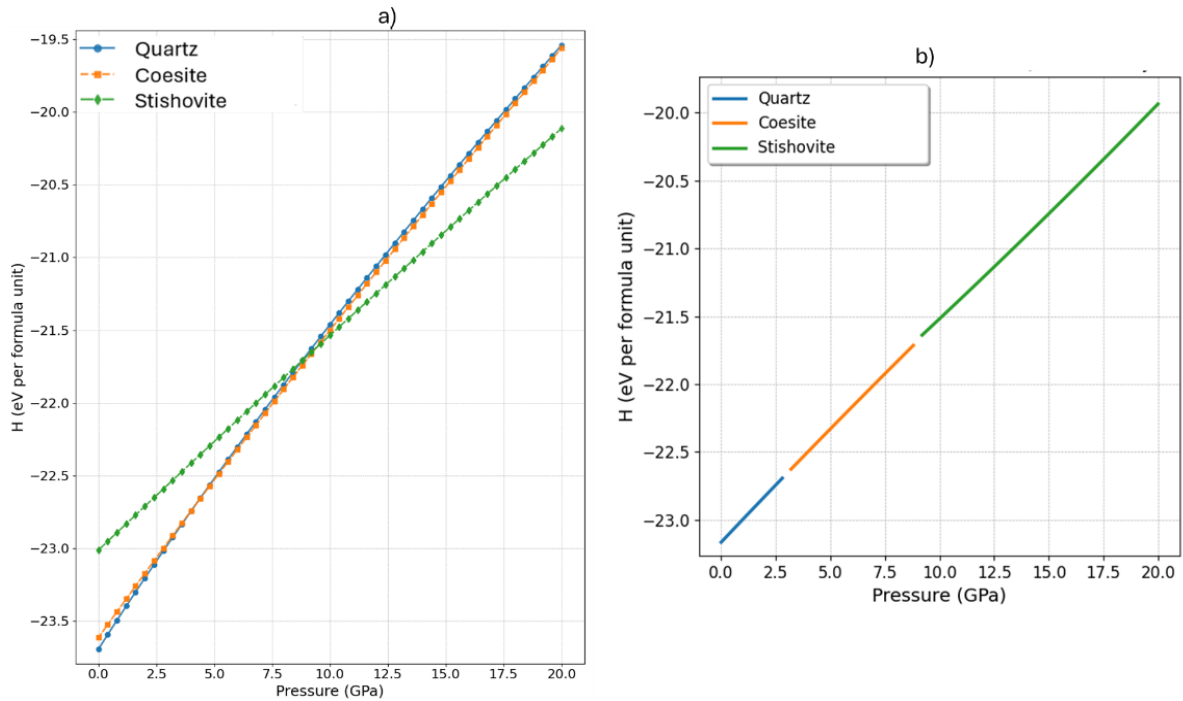

**Figure 5**. (a) shows the variation of enthalpy (H) per formula unit as a function of pressure (GPa) for three polymorphs of silica: quartz, coesite, and stishovite, and (b) illustrates the stability regions of each phase over the pressure range. The figure presents the enthalpy behaviour of each phase under increasing pressure up to 20 GPa.

**Table 3|** Experimental and Calculated Phase Transition Pressures for polymorphs of silica. Experimental values for comparison are taken from ref. [47].

| Property | quartz to coesite | coesite to stishovite |
| --- | --- | --- |
| Exp. Pressure (GPa) | 2-2.5[47] | 8-8.5[47], ~9[25] |
| Calc. Pressure (GPa) | ~3.5 | ~9 |

*Enthalpy as a function of pressure in Silicalite polymorphs*

Purely siliceous silicalite shows polymorphism, crystallizing in an orthorhombic (Pnma)[28], monoclinic (P21/n11)[29], and orthorhombic (P212121)[30] lattice. A monoclinic-to-orthorhombic phase transition (MOPT) has been observed, where the monoclinic phase is stable below transition temperatures, and the orthorhombic phase is stable at above transition temperature[68]. On the other hand, the pressure-induced MOPT was reported to take place between 1 and 1.5 GPa at a constant room temperature [65, 66], while there is still uncertainty in the experimental literature whether it is reversible [66, 69] or not[65] upon decompression. To this end, we have also predicted a phase transition around 1.4 GPa, in very good agreement with already reported data[65, 66]—see below **Figure 6**.

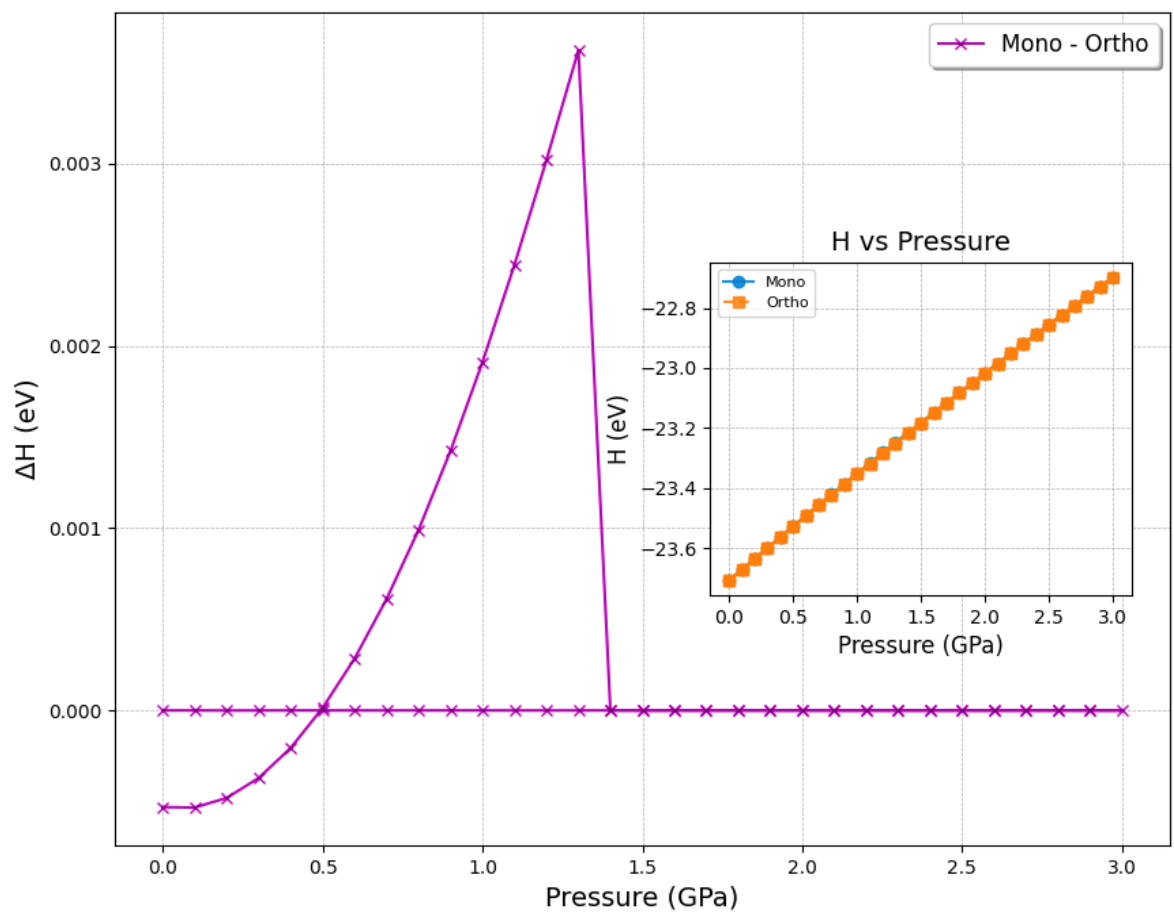

**Figure 6**. The main plot shows the enthalpy difference ($\Delta H$) between the monoclinic and orthorhombic structures of ZSM-5 as a function of pressure (GPa). The inset plot, located at the lower right, displays the enthalpy (H) per formula unit for both the monoclinic and orthorhombic structures across the same pressure range.

*Fluoride Chemistry in Zeolites*

The role of fluoride ions (F⁻) in zeolite synthesis is well-established. Fluoride ions are frequently used as mineralizing agents in hydrothermal synthesis, where they aid the formation of highly crystalline, defect-free silica frameworks[70]. Fluoride can adopt various configurations in Zeolites, as outlined by Attfield *et al.*[70], who identified three primary environments for F⁻ ions: (i) as part of an ion pair near a structure-directing agent (SDA), (ii) centrally located in small cages, and (iii) coordinating with Si to form pentacoordinated $SiO_4F^-$ units. The inclusion of fluoride may help stabilize the double 4-ring *(D4R) structure, as observed in earlier studies by Flanigen and Patton, who first noted the importance of F⁻ in promoting zeolite formation[71]. We examined the similar behaviour of F⁻ within different zeolite frameworks using the MACE ML-IP model as shown in **Figure 7**. Our results show that F⁻, when located inside a D4R, consistently positions itself at the center of the ring (case ii) and does not form direct coordination with any silicon (Si) atoms which is consistent with our earlier DFT reported data[70], where fluoride ions are described as residing in small D4R cages, far from Si atoms. Also, the formation of pentacoordinated Si species, in which F⁻ coordinates directly with Si, has been already reported[72] showing that F⁻ forms part of a trigonal bipyramidal $SiO_4F^-$ unit (case iii) in zeolites like MFI, FER and CHA which agree with our findings.

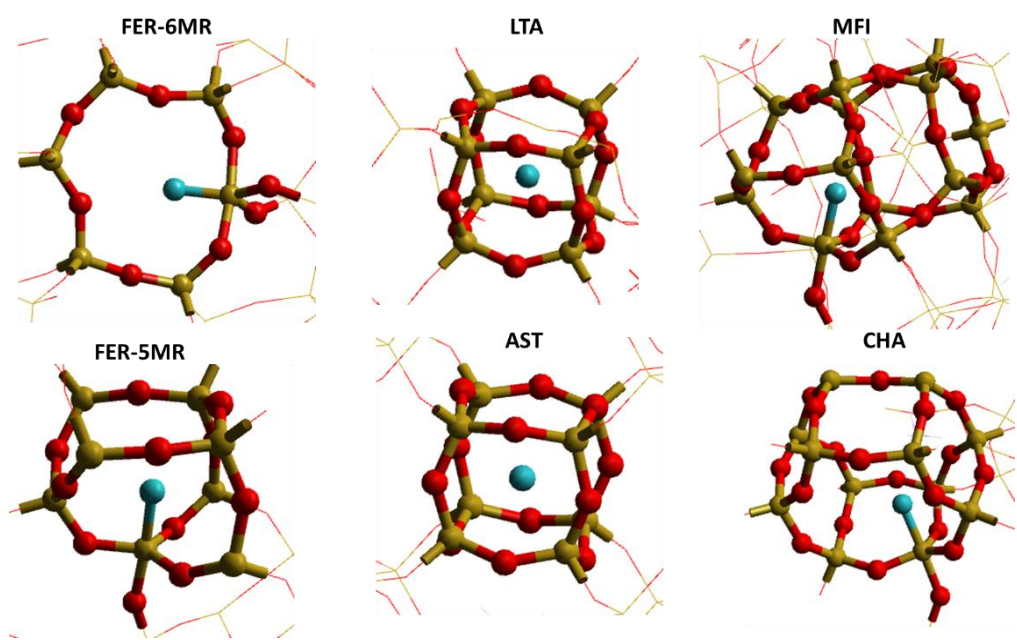

**Figure 7**. A schematic representation of the various types of F⁻ ion environments found in siliceous zeolites: including F⁻ in the center of a small cage of d4R without coordinating to Si atoms (LTA and AST), and F⁻ coordinated to a Si atom to form part of a pentacoordinated $SiO_4F^-$ unit (FER, MFI and CHA). The framework $SiO_2$ is shown using a wire framework motif.

## Summary and Conclusion

We have examined the structural and thermodynamic properties of silica polymorphs and siliceous zeolites using the MACE ML-IP model. The MACE model, which offers near-DFT accuracy with greater computational efficiency, was used to predict framework energies and phase transitions in silica polymorphs. The MACE ML-IP results showed excellent agreement with both DFT+D3 calculations and experimental data, particularly in predicting framework energies for various zeolite structures. In comparison to earlier IP models, MACE ML-IP delivered more accurate results, closely matching DFT predictions for complex frameworks like including MFI/ZSM-5, MEI/ZSM-18, and STT/SSZ-23. The model also accurately predicted phase transitions for silica polymorphs, including quartz to coesite (~3 GPa) and coesite to stishovite (~9 GPa), aligning well with experimental observations. In ZSM-5 polymorphs, the monoclinic-to-orthorhombic transition around 1.35 GPa was observed which also corresponds well with experimental data. Structural changes under pressure were also analysed for both silica and ZSM-5 polymorphs which showed distinct compression patterns. Furthermore, we explored fluoride ions' role in zeolite synthesis, highlighting their stabilizing effects in double four-membered rings and pentacoordinated $SiO_4F^-$ units, consistent with known behaviours in zeolites. Our MACE ML-IP model demonstrates accuracy and efficiency in predicting the energetics and structural responses of zeolites and silica polymorphs, and its close alignment with experimental data and DFT predictions makes it a valuable tool in zeolite research, evaluating new frameworks and understanding phase transitions.

## Acknowledgements

The research utilised the STFC Scientific Computing Department's SCARF cluster and the Thomas Young Tier 2 HPC platform. AME acknowledge EPSRC grant EP/V028537/1. We also acknowledge EP/X035859/1, EP/W026775/1 and EP/T022213/1.